# Nucleophilicity/ Electrophilicity Excess in Analyzing Molecular Electronics

**D. R. Roy[1], V. Subramanian[2,*] and P. K. Chattaraj[1,*]**

[1]*Department of Chemistry, Indian Institute of Technology, Kharagpur 721302, INDIA*

[2]*Chemical Laboratory, Central Leather Research Institute, Adyar, Chennai 600 020, INDIA*

**Abstract:**

Intramolecular electron transfer capability of all metal aromatic and anti-aromatic aluminum cluster compounds is studied in terms of density functional theory based global and local reactivity descriptors. This study will provide important inputs towards the fabrication of the material required for molecular electronics.

[*]Authors for correspondence: pkc@chem.iitkgp.ernet.in, subuchem@hotmail.com

## 1. Introduction

The concept of aromaticity is extended to the all-metal molecules in the very recent past.[1–9] The area of this specialization is carefully termed as 'metallaaromaticity'.[1–9] The investigation on aromaticity of the various ionic units of Al, Ga, In, Hg, Si etc. and their neutral and ionic complexes is one of the most interesting research topics in the recent literature.[1–9] Both the experimental like laser vaporization technique using photoelectron spectroscopy[1,2] and ab-initio or density functional theory (DFT)[10,11] based analysis are performed to characterize the nature of those ionic units and their complexes. All-metal aromatic compounds, viz., $MAl_4^-$ (M = Li, Na, K and Cu) are synthesized by Li et al[1] for the first time. The square planar geometry and the presence of two delocalized π- electrons in the $Al_4^{2-}$ dianion makes it aromatic by obeying Hückel's (4n+2) π-electron rule. The transformation of a nonaromatic $Al_4Cl_4(NH_3)_4$ molecule into a π-aromatic $Na_2Al_4Cl_4(NH_3)_4$ molecule have also shown by theoretical investigation.[3] Chattaraj et al[4] have proposed two new aromaticity indices based on the polarizability (α) and hardness (η) from the electronic structure principles of Density functional theory (DFT). [12-21]

All-metal antiaromatic molecule $Li_3Al_4^-$ is synthesized by Kuznetsov et al for the first time.[2] The presence of four π- electrons obeying Hückel's 4n rule and the rectangular structure of its $Al_4^{4-}$ tetraanion unit provides antiaromatic nature of that molecule. It is shown that $Al_4^{4-}$ is overall antiaromatic through the electron localization function (ELF) analysis.[5] On the other hand these molecules have been shown be to net aromatic[6,7] according to the magnetic criterion of aromaticity like nucleus independent chemical shift[6] and magnetic field induced current density[7] analysis because σ-aromaticity overwhelms its π- antiaromaticity. This controversy[22] is still a point of interest of the recent literature.[1-9]

The important insights into the reactivity and electronic properties[24-30] of these multi-metallic clusters are obtained through various aspects of alloy formation.[24-30] Application of the aluminum alloys is spread over the electronic, mechanical and optical devices, corrosion protection, and aerospace engineering etc. for the last two decades. The recent trends on the development and application of the metallurgy of aluminum powders and alloys are of great interest.[24] The application of Al alloy brazing sheet in automobile heat exchangers is also very useful.[25] The Al alloys are used as a cathode in corrosion protection.[26] With the structure of tris(8-hydroxy quinoline)aluminum ($Alq^3$) Kim et al[27] fabricated and characterized double-layer-type electroluminescent devices. The aluminum-copper-lithium alloys have a wide application in the field of aerospace engineering.[28]

Park et al[29] have experimentally studied the charge transfer effect in aluminum-magnesium alloy formation. They showed that charge transfer from Mg to Al takes place on alloying. Chang et al[30] have investigated the effect of charge transfer and hybridization in $Ni_3Al$ and $Ni_3Ga$ alloys with the X-ray absorption spectroscopy and theoretical calculations. They found that Al losses some p-orbital charge whereas Ni gains some charge to form $Ni_3Al$. As a consequence the study of the intra and inter cluster reactivity of the all-metal aromatic and anti-aromatic molecules in terms the DFT based local reactivity descriptors, viz., fukui function (FF) [10] and philicity[15] has gained importance as no such work in detail has been done on this issue.

Density functional theory[10,11] based global reactivity descriptors (e.g., electronegativity,[12, 13] chemical potential,[13] hardness[14,15] and electrophilicity[16]) are quite successful in unraveling chemical reactivity. Fukui function (FF)[10] is one of the widely used local density functional descriptors to model chemical reactivity and site selectivity. The atom with highest FF is highly reactive compared to the other atoms in the molecule. The FF is defined as the derivative of the electron density $\rho(\vec{r})$ with respect to the total number of electrons $N$ in the system, at constant external potential $v(\vec{r})$ acting on an electron due to all the nuclei in the system[10]

$$f(\vec{r}) = \left[\delta\mu/\delta v(\vec{r})\right]_N = \left[\partial\rho(\vec{r})/\partial N\right]_{v(\vec{r})} \quad (1)$$

where $\mu$ is the chemical potential of the system.

Parr et al[10] defined the elecrophilicity index ($\omega$), which measures the stabilization in energy when a system acquires an additional electronic charge from the environment, as follows:

$$\omega = \mu^2 / 2\eta \quad (2)$$

where μ and η are the chemical potential and hardness respectively.

The generalized concept of philicity was proposed by Chattaraj et al[17] which contains almost all information about hitherto known different global and local reactivity and selectivity descriptors, in addition to the information regarding electrophilic/nucleophilic power of a given atomic site in a molecule. It is possible to define a local quantity called philicity associated with a site k in a molecule with the aid of the corresponding condensed- to- atom variants of Fukui function $f_k^\alpha$ as,[17]

$$\omega_k^\alpha = \omega f_k^\alpha \quad (3)$$

where ($\alpha$ = +, - and 0) represents local philic quantities describing nucleophilic, electrophilic and radical attacks. Eq. (3) predicts that the most electrophilic site in a molecule is the one providing the maximum value of $\omega_k^+$. This site also coincides with the softest site in a molecule. When two molecules react, which one will act as an electrophile (nucleophile) will depend on, one which has a higher (lower) electrophilicity index. This global trend originates from the local behavior of the molecules or precisely the atomic site (s) that is (are) prone to electrophilic (nucleophilic) attack**.** Chattaraj *et al* established a generalized treatment of both global and local electrophilicity, as well as nucleophilicity.

The group concept of philicity is very useful in unraveling reactivity of various molecular systems.[18] The condensed philicity summed over a group of relevant atoms is defined as the "group philicity". It can be expressed as

$$\omega_g^\alpha = \sum_{k=1}^{n} \omega_k^\alpha \quad (4)$$

where n is the number of atoms coordinated to the reactive atom, $\omega_k^\alpha$ is the local electrophilicity of the atom k, and $\omega_g^\alpha$ is the group philicity obtained by adding the local philicity of the nearby bonded atoms, where ($\alpha$= +, -, 0) represents nucleophilic, electrophilic and radical attacks respectively.

The purpose of the present study is to investigate the intramolecular reactivity of the aromatic $Al_4^{2-}$ dianion and anti-aromatic $Al_4^{4-}$ tetraanion units associated with various all-metal complexes and also the intermolecular reactivity of those units among the molecules of different categories. The stable isomers of aromatic $Al_4^{2-}$ and $MAl_4^-$ (M=Li, Na, K and Cu) (Figures 1, 2) and anti-aromatic $Al_4^{4-}$, $Li_3Al_4^-$ and $Li_4Al_4$ isomers[4] (Figures 3-5) are chosen in the present study. Also the sandwich complexes based on aromatic molecules[8] (Figure 2) and anti-aromatic molecules[9] (Figure 6) are selected. Section 2 defines two important molecular electronics descriptors for this purpose. Theoretical background is provided in section 3 and the computational details are presented in section 4. Section 5 presents the results and discussion and finally some concluding remarks are given in section 6.

## 2. Nucleophilicity/ Electrophilicity Excess

Based on the group concept, for a nucleophile in a molecule should have to posses more group philicity due to electrophilic attack over nucleophilic attack on it. This difference can be expressed with a nucleophilicity excess ($\Delta\omega_g^{\mp}$) along the line of dual descriptor[31] as

$$\Delta\omega_g^{\mp} = \omega_g^- - \omega_g^+ = \omega\left(f_g^- - f_g^+\right) \tag{5}$$

where $\omega_g^- (\equiv \sum_{k=1}^{n} \omega_k^-)$ and $\omega_g^+ (\equiv \sum_{k=1}^{n} \omega_k^+)$ are the group philicities of the nucleophile in the molecule due to electrophilic and nucleophilic attack respectively. It is expected that the nucleophilicity excess ($\Delta\omega_g^{\mp}$) for a nucleophile should always be positive whereas it will provide a negative value for an electrophile in a molecule.

Corresponding electrophilicity excess ($\Delta\omega_g^{\pm}$) for an electrophile in a molecule should posses more group philicity due to nucleophilic attack over the electrophilic attack on it. The electrophilicity excess ($\Delta\omega_g^{\pm}$) for an electrophile can be expressed as

$$\Delta\omega_g^{\pm} = -\Delta\omega_g^{\mp} = \omega_g^+ - \omega_g^- = \omega\left(f_g^+ - f_g^-\right) \tag{6}$$

where $\omega_g^+$ and $\omega_g^-$ are the group philicities of the electrophile in the molecule due to nucleophilic and electrophilic attack respectively. It is expected that the electrophilicity excess ($\Delta\omega_g^{\pm}$) for an electrophille should always be positive whereas it will provide a negative value for a nucleophile in a molecule.

For a molecular system with only two distinct units, the nucleophilicity excess ($\Delta\omega_g^{\mp}$) of the nucleophile should be equal to the electrophilicity excess ($\Delta\omega_g^{\pm}$) of the electrophille, as expected from the conservation of FF and philicity, i.e.

$$\Delta\omega_g^{\mp} \text{ (nucleophile)} = \Delta\omega_g^{\pm} \text{ (electrophile)} \tag{7}$$

## 3. Theoretical Background

The quantitative definitions for chemical potential (μ) and electronegativity (χ)[12,13] for an N– electron system with total energy E can respectively be given as

$$\mu = \left[\frac{\partial E}{\partial N}\right]_{v(\vec{r})} \tag{8}$$

and

$$\chi = -\mu = -\left(\frac{\partial E}{\partial N}\right)_{v(\vec{r})} \tag{9}$$

where $v(\mathrm{r})$ is the external potential.

Chemical hardness (η) has been identified as an useful global reactivity index in atoms, molecules and clusters.[14,15] The theoretical definition of chemical hardness has been provided by DFT as the second derivative of electronic energy with respect to the number of electrons (N), for a constant external potential $v(\mathrm{r})$, viz.,

$$\eta = \frac{1}{2}\left(\frac{\partial^2 E}{\partial N^2}\right)_{v(\vec{r})} = \frac{1}{2}\left(\frac{\partial \mu}{\partial N}\right)_{v(\vec{r})} \tag{10}$$

Using a finite difference method the working equations for the calculation of chemical potential, electronegativity and chemical hardness can be given by

$$\mu = -\frac{IP + EA}{2} \;\; ; \;\; \chi = \frac{IP + EA}{2} \;\; ; \;\; \eta = \frac{IP - EA}{2} \tag{11}$$

where *IP* and *EA* are ionization potential and electron affinity of the system respectively.

Using the ΔSCF finite difference approach the *IP* and *EA* can be calculated for the *N*-electron system as follows:

$$IP \approx E\,(N\text{-}1) - E\,(N) \qquad ; \qquad EA \approx E\,(N) - E\,(N\text{+}1) \tag{12}$$

where $E\,(N)$ is the electronic energy for the *N* electron system.

Depending on the electron transfer, three types of FF[10] are defined as

$$f^{+}(\vec{r}) = \rho_{N+1}(\vec{r}) - \rho_{N}(\vec{r}) \quad \text{for nucleophilic attack} \quad (13a)$$

$$f^{-}(\vec{r}) = \rho_{N}(\vec{r}) - \rho_{N-1}(\vec{r}) \quad \text{for electrophilic attack} \quad (13b)$$

$$f^{0}(\vec{r}) = [\rho_{N+1}(\vec{r}) - \rho_{N-1}(\vec{r})]/2 \quad \text{for radical attack} \quad (13c)$$

The condensed FF are calculated using the procedure proposed by Yang and Mortier[23] based on a finite difference method

$$f_k^{+} = q_k(N+1) - q_k(N) \quad \text{for nucleophilic attack} \quad (14a)$$

$$f_k^{-} = q_k(N) - q_k(N-1) \quad \text{for electrophilic attack} \quad (14b)$$

$$f_k^{o} = [q_k(N+1) - q_k(N-1)]/2 \quad \text{for radical attack} \quad (14c)$$

where $q_k$ is the electronic population of atom k in a molecule.

The electric dipole polarizability is a measure of the linear response of the electron density in the presence of an infinitesimal electric field F and it represents a second order variation in energy

$$\alpha_{a,b} = -\left(\frac{\partial^2 E}{\partial F_a \partial F_b}\right) \quad a,b = x,y,z \quad (15)$$

The polarizability $\alpha$ is calculated as the mean value as given in the following equation

$$\langle\alpha\rangle = \tfrac{1}{3}\left(\alpha_{xx} + \alpha_{yy} + \alpha_{zz}\right) \quad (16)$$

## 4. Computational details

All the all-metal aromatic and anti-aromatic molecules, viz., $Al_4^{2-}$, $MAl_4^{-}$ (M=Li, Na, K and Cu), $Al_4^{4-}$, $Li_3Al_4^{-}$, $Li_4Al_4$ are minimized in the B3LYP method with the 6-311+G* basis set. To optimize the aromatic sandwich complexes[8] $M_2(Al_4TiAl_4)$ (M=Li, Na and K) B3LYP/6-311G level of calculation is followed. Also for anti-aromatic sandwich complexes[9] $Li_4Al_4Fe(CO)_3$, $(Li_4Al_4)_2Ni$ and bis($Li_4Al_4$ nickel(II) chloride) single point calculation with the same level of theory, B3LYP/6-311G**, is used with the geometry as reported in reference.[9] For calculation of (N+1) and (N-1) systems same geometry of the N-electron system is used. Using ΔSCF method, the ionization potential (IP) and electron affinity (EA) are calculated using equation (12). The $\eta$, $\mu$ and $\alpha$ are calculated using equations (11, 16). The electrophilicity index ($\omega$) is calculated using equation (2). The fukui function (FF), philicities are calculated using equations (14, 3, 4).

## 5. Results and discussion

A careful study on the electronic structure, property and reactivity of all-metal aromatic compounds, viz., $Al_4^{2-}$, $MAl_4^{-}$ (M=Li, Na, K and Cu) and aromatic sandwich complexes $M_2(Al_4TiAl_4)$ (M=Li, Na, K) and anti-aromatic compounds, viz., $Al_4^{4-}$, $Li_3Al_4^{-}$, $Li_4Al_4$ and anti-aromatic sandwich complexes $Li_4Al_4Fe(CO)_3$, $(Li_4Al_4)_2Ni$ and bis($Li_4Al_4$ nickel(II) chloride) has been made. It can be noticed that in all the molecules, four membered aluminum unit $Al_4$ is present and it may be considered as a superatom.[32] This super unit can easily take part in charge transfer process with the M (≡Li, Na, K, Cu, Fe, Ni, Ti) atom in those all-metal aromatic and anti-aromatic complexes.

### *5.1 Electronic properties and reactivity of all-metal aromatic compounds*

Table 1 presents global electronic properties, e.g. energy (E), polarizability ($\alpha$), chemical hardness ($\eta$), chemical potential ($\mu$) and the electrophilicity ($\omega$) of all-metal aromatic compounds, viz., $Al_4^{2-}$, $MAl_4^{-}$ (M=Li, Na, K and Cu) and aromatic sandwich complexes $M_2(Al_4TiAl_4)$ (M=Li, Na, K) (Figures 1, 2). The $D_{4h}$ isomer of $Al_4^{2-}$ (Figure 1) and $C_{4v}$ isomer of $MAl_4^{-}$ (Figure 2) are energetically most stable.[1,4] Also the energetically most stable isomer of aromatic molecules is found to be the hardest and the least polarizable.[4]

Table 2 shows the fukui function (FF) and philicity values at the each atomic site of $Al_4^{2-}$ isomers. It is found that in the case of $D_{4h}$ isomer the FF ($f^+$, $f^-$) and philicity ($\omega^+$, $\omega^-$) values for (nucleo/electro) philic attack at each aluminum site are almost equal, as expected in this square planar structure. For the other two stable isomers ($D_{\infty h}$ and $D_{3h}$) the FF as well as philicity values at each atomic site are not equal due to the absence of symmetric electron localization.

Table (3-6) presents the group fukui function ($f_g^+$, $f_g^-$) and group philicity ($\omega_g^+$, $\omega_g^-$) values of the $Al_4^{2-}$ nucleophile and $M^+$ (M=Li, Na, K, Cu) electrophile in the $MAl_4^-$ isomers. It is found that in all $MAl_4^-$ isomers the nucleophilicity of the $Al_4^{2-}$ aromatic super atom[32] overwhelms its electrophilic trend (i.e. $f_g^- \succ f_g^+$ and $\omega_g^- \succ \omega_g^+$) and therefore $\Delta\omega_g^{\mp}$ is positive, whereas the electrophilicity of $M^+$ dominates over its nucleophilicity (i.e. $f_g^+ \succ f_g^-$ and $\omega_g^+ \succ \omega_g^-$) and therefore $\Delta\omega_g^{\mp}$ is negative as expected. It is important to note that $\Delta\omega_g^{\mp}$ of $Al_4^{2-}$ is maximum in the case of most stable $C_{4v}$ isomer of the $MAl_4^-$ molecule. The order of the $\Delta\omega_g^{\mp}$ value of $Al_4^{2-}$ nucleophile in $MAl_4^-$, $C_{4v} \succ C_{2v} \succ C_{\infty v}$, i.e. stabilization of an $MAl_4^-$ isomer (except in $KAl_4^-$) increases its nucleophilicity and accordingly can be used as a better molecular cathode. The group fukui function and group philicity values of the aromatic sandwich complexes $M_2(Al_4TiAl_4)$ are reported in Table 7. It may be noted that the group nucleophilicity of the $Al_4^{2-}$ unit in those complexes dominates over its group electrophilicity, i.e. $\Delta\omega_g^{\mp}$ is positive as expected. It is also important to note that the nucleophilicity of the $Al_4^{2-}$ unit in $MAl_4^-$ ($C_{4v}$) increases as $Cu \prec K \prec Na \prec Li$ whereas the order gets reversed for $M_2(Al_4TiAl_4)$ sandwich complexes as dictated by the respective nucleophilicity excess values.

## *5.2. Electronic properties and reactivity of all-metal anti-aromatic compounds*

Table 8 represents the global electronic properties, viz., energy (E), polarizability (α), hardness (η), chemical potential (μ) and the electrophilicity (ω) of the various isomers of all-metal anti-aromatic compounds, viz., $Al_4^{4-}$ (singlet and triplet), $Li_3Al_4^-$, $Li_4Al_4$ (Figures 3-5) and their sandwich complexes $Li_4Al_4Fe(CO)_3$, $(Li_4Al_4)_2Ni$ and bis($Li_4Al_4$ nickel(II) chloride) (Figure 6). The linear $Al_4^{4-}$ is found to be slightly more stable[4] compared to its cyclic counterpart in both the singlet and triplet states. The $C_S$ (Fork) isomer of $Li_3Al_4^-$ is found to be energetically the most stable, the least polarizable and the hardest.[2, 4] The $C_{2h}$ isomer of $Li_4Al_4$ is the most stable.

Tables (9, 10) present the fukui function and philicity values at each atomic site of $Al_4^{4-}$ isomers (Figure 3) of both singlet and triplet states. The unequal Fukui function ($f^+$, $f^-$) and philicity ($\omega^+$, $\omega^-$) values may be viewed as an effect of localized electrons. Two Al atoms behave differently than the remaining two vindicating the rectangular (not square) geometry of $Al_4^{4-}$ ion.

Table 11 presents the group fukui function ($f_g^+, f_g^-$) and group philicity ($\omega_g^+$, $\omega_g^-$) values of all the $Li_3Al_4^-$ isomers (Figure 4) for nucleophilic and electrophilic attacks respectively. The positive $\Delta\omega_g^{\mp}$ values of the $Al_4^{4-}$ unit in all $Li_3Al_4^-$ isomers provide the nucleophilic nature of $Al_4^{4-}$ unit in those compounds. Also negative $\Delta\omega_g^{\mp}$ values of $Li_3^{3+}$ unit in all isomers of $Li_3Al_4^-$ imply its electrophilic nature over its nucleophilic trend.

Table 12 shows the group Fukui function ($f_g^+, f_g^-$) and group philicity ($\omega_g^+$, $\omega_g^-$) values of all the $Li_4Al_4$ isomers (Figure 5) for nucleophilic and electrophilic attacks respectively. The positive $\Delta\omega_g^{\mp}$ values of the $Al_4^{4-}$ unit in all $Li_4Al_4$ isomers provide the nucleophilic nature of $Al_4^{4-}$ unit in those compounds. Also negative $\Delta\omega_g^{\mp}$ values of $Li_4^{4+}$ unit in all isomers of $Li_4Al_4$ imply its electrophilic nature over nucleophilic trend.

We have also investigated the nucleophilicity of the antiaromatic $Al_4^{4-}$ unit in the anti-aromatic sandwich complexes.[9] $Li_4Al_4Fe(CO)_3$, $(Li_4Al_4)_2Ni$ and bis($Li_4Al_4$ nickel(II) chloride) (Figure 6). The positive $\Delta f_g^{\mp}$ and $\Delta\omega_g^{\mp}$ values $Al_4^{4-}$ units in those sandwich complexes provides its nucleophilic nature.

## 6. Conclusion

It has been demonstrated through the analysis of nucleophilicity/electrophilicity excess values of all-metal aromatic and anti-aromatic cluster compounds and their alkali metal and sandwich complexes that the $Al_4^{2-}$ /$Al_4^{4-}$ unit behaves as a nucleophile in all cases wherein the electrophiles will prefer to attack. Important insights into the associated molecular electronics can be obtained through this systematic study of electron localization pattern in the $Al_4$ group by changing the attached metal ion.

**Acknowledgment**

We thank CSIR, New Delhi for financial assistance.

References

1. Li, X.; Kuznetsov, A. E.; Zhang, H.-F.; Boldyrev, A. I.; Wang, L.-S. *Science* 2001, *291*, 859.
2. Kuznetsov, A.; Birch, K.; Boldyrev, A. I.; Li, X.; Zhai, H.; Wang, L.-S. *Science* 2003, *300*, 622.
3. Kuznetsov, A. E.; Zhai, H. J.; Wang, L.-S.; Boldyrev, A. I. *Inorg. Chem. Commn.* 2002, *41*, 6062.
4. Chattaraj, P. K.; Roy, D. R.; Elango, M.; Subramanian, V. *J. Phys. Chem. A* 2005, *109*, 9590.
5. Santos, J. C.; Andres, J.; Aizman, A.; Fuentealba, P. *J. Chem. Theo. Comp.* 2005, *1*, 83.
6. Schleyer, P. v. R.; Maerker, C.; Dransfeld, A.; Jiao, H.; Hommes, N. J. R. v. E. *J. Am. Chem. Soc.* 1996, *118*, 6317; Schleyer, P. v. R.; Jiao, H.; Hommes, N. v. E.; Malkin, V. G.; Malkina, O. L. *J. Am. Chem. Soc.* 1997, *119*, 12669. Chen, Z.; Corminboeuf, C.; Heine, T.; Bohmann, J.; Schleyer, P. v. R. *J. Am. Chem. Soc.* 2003, *125*, 13930.
7. Havenith, R. W. A.; Fowler, P. W.; Steiner, E.; Shetty, S.; Kanhere, D.; Pal, S. *Phys. Chem. Chem. Phys.* 2004, *6*, 285.
8. Mercero, J. M.; Ugalde, J. M. *J. Am. Chem. Soc.* 2004, *126*, 3380.
9. Datta, A.; Pati, S. K. *J. Am. Chem. Soc.* 2005, *127*, 3496.
10. Parr, R. G.; Yang, W. *Density Functional Theory of Atoms and Molecules*; Oxford University Press: New York, 1989.
11. Geerlings, P.; De Proft, F.; Langenaeker, W. *Chem.Rev.* 2003, *103*, 1793.
12. Pauling, L. *The Nature of the Chemical Bond*, 3rd ed.; Cornell University Press: Ithaca, NY, 1960.
13. Parr, R. G.; Donnelly, R. A.; Levy, M.; Palke, W. E. *J. Chem. Phys.* 1978, *68*, 3801.
14. Pearson, R. G. *J. Am. Chem. Soc.* 1963, *85*, 3533.
15. Parr, R. G.; Pearson, R. G. *J. Am. Chem. Soc.* 1983, *105*, 7512. Chattaraj, P. K.; Lee, H.; Parr, R. G. *J. Am. Chem. Soc.* 1991, *113*, 1855. Chattaraj, P. K.;

Schleyer, P. v. R. *J. Am. Chem. Soc*. 1994, *116*, 1067. Chattaraj, P. K.; Maiti, B. *J. Am. Chem. Soc.* 2003, *125*, 2705.

16. Parr, R. G.; Szentpaly, L. v.; Liu, S. *J. Am. Chem. Soc.* 1999, *121*, 1922.
17. Chattaraj, P. K.; Maiti, B.; Sarkar, U. *J. Phys. Chem. A*, **2003**, 107**,** 4973.
18. Parthasarathi, R.; Padmanabhan, J.; Elango, M.; Subramanian, V.; Chattaraj, P. K. *Chem. Phys. Lett.* 2004, *394*, 225.
19. Pearson, R. G. *J. Chem. Educ.* 1987, *64*, 561. Parr, R. G.; Chattaraj, P. K. *J. Am. Chem. Soc.* 1991, *113*, 1854. Ayers, P. W.; Parr, R. G. *J. Am. Chem. Soc.* 2000, *122*, 2010. Chattaraj, P. K.; Fuentealba, P.; Gomez, B.; Contreras, R. *J. Am. Chem. Soc*. 2000, *122*, 348.
20. Chattaraj, P. K.; Sengupta, S. *J. Phys. Chem.* 1996, *100*, 16126.
21. Ghanty, T. K.; Ghosh, S. K. *J. Phys. Chem.* 1996, *100*, 12295.
22. Ritter, S. *Chem. Eng. News* 2003, *81*, 23.
23. Yang, W.; Mortier, W. J. *J. Am. Chem. Soc*. 1986, *108*, 5708.
24. Stojadinovic, S.; Vobornik, S.; Gulisija, Z. *Tehnika (Belgrade)* 1994, 49, RGM1.
25. Su, J.; Horng, W. *Jishu Yu Xunlian* 2002, 27, 19.
26. Umino, T. *Bosei Kanri* 1995, 39, 286.
27. Kim, J-Y.; Kim, E-S.; Choi, J-H. *J. Appl. Phys.* 2002, 91, 1944.
28. Sverdlin, A.; Drits, A.; Ovchinnikov, V. *Proce. Mater. Solutions Conf.* (*Indianapolis, IN*), *U.S.A.*, Nov. 5-8, 2001, 255.
29. Park, C. Y.; Sagawa, T. *J. Appl. Phys.* 1986, 60, 1310.
30. Chang, Y. K.; Lin, K. P.; Pong, W. F.; Tsai, M.-H.; Hseih, H. H.; Pieh, J. Y.; Tseng, P. K.; Lee, J. F.; Hsu, L. S. *J. Appl. Phys.* 2000, 87, 1312.
31. Morell, C.; Grand, A.; Toro-Labbe´, A. *J. Phys. Chem. A* 2005, *109*, 205.
32. Chako, S.; Deshpande, M.; Kanhere, D. G. *Phys. Rev. B.* 2001, *64*, 155409. Kumar, V. *Phys. Rev. B* 1999, *60*, 2916.

***Table 1.*** Energy (E), Polarizability (α), Hardness (η), Chemical potential (μ) and the Electrophilicity (ω) values of Different Isomers of $Al_4^{2-}$ and $MAl_4^-$ and Sandwich Complexes.

| Molecule | PG | Energy (E) | $\alpha$ | $\eta$ | $\mu$ | $\omega$ |
|---|---|---|---|---|---|---|
| *$Al_4^{2-}$ Isomers* | | | | | | |
| $Al_4^{2-}$ | $D_{\infty h}$ | -969.698 | 675.819 | 1.512 | 3.106 | 3.191 |
| | $D_{3h}$ | -969.702 | 665.292 | 1.741 | 3.415 | 3.349 |
| | $D_{4h}$ | -969.741 | 525.790 | 1.965 | 3.625 | 3.344 |
| *$MAl_4^-$ Isomers* | | | | | | |
| $LiAl_4^-$ | $C_{\infty v}$ | -977.278 | 686.377 | 1.273 | -0.183 | 0.013 |
| | $C_{2v}$ | -977.357 | 414.451 | 1.765 | -0.155 | 0.007 |
| | $C_{4v}$ | -977.362 | 375.031 | 3.622 | 1.633 | 0.368 |
| $NaAl_4^-$ | $C_{\infty v}$ | -1132.063 | 719.792 | 1.272 | -0.192 | 0.014 |
| | $C_{2v}$ | -1132.135 | 458.250 | 1.671 | -0.178 | 0.009 |
| | $C_{4v}$ | -1132.144 | 404.265 | 2.672 | 0.810 | 0.123 |
| $KAl_4^-$ | $C_{\infty v}$ | -1569.696 | 922.935 | 1.090 | -0.181 | 0.015 |
| | $C_{2v}$ | -1569.768 | 574.811 | 1.438 | -0.201 | 0.014 |
| | $C_{4v}$ | -1569.777 | 471.425 | 1.545 | -0.164 | 0.009 |
| $CuAl_4^-$ | $C_{\infty v}$ | -2610.285 | 500.603 | 1.662 | -0.122 | 0.004 |
| | $C_{2v}$ | -2610.363 | 343.957 | 2.370 | 0.145 | 0.004 |
| | $C_{4v}$ | -2610.371 | 331.243 | 2.933 | -0.426 | 0.031 |
| *Aromatic Sandwich Complexes* | | | | | | |
| $Li_2(Al_4TiAl_4)$ | | -2804.008 | 594.629 | 1.577 | -3.422 | 3.713 |
| $Na_2(Al_4TiAl_4)$ | | -3113.572 | 654.603 | 1.519 | -3.266 | 3.510 |
| $K_2(Al_4TiAl_4)$ | | -3988.850 | 738.307 | 1.438 | -2.917 | 2.959 |

***Table 2.*** Fukui function ($f^+$, $f^-$) and Philicity ($\omega^+, \omega^-$) Values for Nucleophilic and Electrophilic Attacks Respectively for the Aluminum Atoms of Different Isomers of $Al_4^{2-}$.

| Isomers | Atom | $f^+$ | $f^-$ | $\omega^+$ | $\omega^-$ |
|---|---|---|---|---|---|
| $Al_4^{2-}$ ($D_{\infty h}$) | Al | 0.4853 | 0.3589 | 1.5483 | 1.1451 |
| | Al | 0.0149 | 0.1403 | 0.0476 | 0.4476 |
| | Al | 0.0151 | 0.1397 | 0.0480 | 0.4458 |
| | Al | 0.4848 | 0.3611 | 1.5467 | 1.1520 |
| $Al_4^{2-}$ ($D_{3h}$) | Al | 0.2258 | 0.3655 | 0.7564 | 1.2243 |
| | Al | 0.0904 | -0.0730 | 0.3027 | -0.2430 |
| | Al | 0.4570 | 0.3376 | 1.5306 | 1.1307 |
| | Al | 0.2268 | 0.3695 | 0.7598 | 1.2375 |
| $Al_4^{2-}$ ($D_{4h}$) | Al | 0.2498 | 0.2496 | 0.8352 | 0.8346 |
| | Al | 0.2510 | 0.2502 | 0.8392 | 0.8366 |
| | Al | 0.2499 | 0.2506 | 0.8357 | 0.8379 |
| | Al | 0.2493 | 0.2496 | 0.8337 | 0.8347 |

***Table 3.*** Group Fukui Function ($f_g^+, f_g^-$) and Group Philicity ($\omega_g^+, \omega_g^-$) Values for Nucleophilic and Electrophilic Attacks Respectively for the Ionic Units of Different Isomers of $LiAl_4^-$.

| Isomers | Ionic Unit | $f_g^+$ | $f_g^-$ | $\Delta f_g^{\mp}$ | $\omega_g^+$ | $\omega_g^-$ | $\Delta\omega_g^{\mp}$ |
|---|---|---|---|---|---|---|---|
| $LiAl_4^-$ ($C_{\infty v}$) | $Al_4^{2-}$ | 0.5262 | 0.7172 | 0.1909 | 0.0070 | 0.0095 | 0.0025 |
| | $Li^+$ | 0.4738 | 0.2828 | -0.1909 | 0.0063 | 0.0037 | -0.0025 |
| $LiAl_4^-$ ($C_{2v}$) | $Al_4^{2-}$ | 0.0019 | 0.8089 | 0.8070 | 1.3E-05 | 0.0055 | 0.0055 |
| | $Li^+$ | 0.9981 | 0.1911 | -0.8070 | 0.0068 | 0.0013 | -0.0055 |
| $LiAl_4^-$ ($C_{4v}$) | $Al_4^{2-}$ | -0.1011 | 0.8052 | 0.9062 | -0.0372 | 0.2965 | 0.3338 |
| | $Li^+$ | 1.1011 | 0.1948 | -0.9062 | 0.4055 | 0.0718 | -0.3338 |

***Table 4.*** Group Fukui Function ($f_g^+, f_g^-$) and Group Philicity ($\omega_g^+, \omega_g^-$) Values for Nucleophilic and Electrophilic Attacks Respectively for the Ionic Units of Different Isomers of $NaAl_4^-$.

| Isomers | Ionic Unit | $f_g^+$ | $f_g^-$ | $\Delta f_g^{\mp}$ | $\omega_g^+$ | $\omega_g^-$ | $\Delta\omega_g^{\mp}$ |
|---|---|---|---|---|---|---|---|
| $NaAl_4^-$ ($C_{\infty v}$) | $Al_4^{2-}$ | 0.4864 | 0.7099 | 0.2235 | 0.0070 | 0.0102 | 0.0032 |
| | $Na^+$ | 0.5136 | 0.2901 | -0.2235 | 0.0074 | 0.0042 | -0.0032 |
| $NaAl_4^-$ ($C_{2v}$) | $Al_4^{2-}$ | -0.0128 | 0.8227 | 0.8356 | -0.0001 | 0.0078 | 0.0079 |
| | $Na^+$ | 1.0128 | 0.1773 | -0.8356 | 0.0096 | 0.0017 | -0.0079 |
| $NaAl_4^-$ ($C_{4v}$) | $Al_4^{2-}$ | -0.0592 | 0.8339 | 0.8930 | -0.0073 | 0.1024 | 0.1097 |
| | $Na^+$ | 1.0592 | 0.1661 | -0.8930 | 0.1301 | 0.0204 | -0.1097 |

***Table 5.*** Group Fukui Function ($f_g^+, f_g^-$) and Group Philicity ($\omega_g^+, \omega_g^-$) Values for Nucleophilic and Electrophilic Attacks Respectively for the Ionic Units of Different Isomers of $KAl_4^-$.

| Isomers | Ionic Unit | $f_g^+$ | $f_g^-$ | $\Delta f_g^{\mp}$ | $\omega_g^+$ | $\omega_g^-$ | $\Delta\omega_g^{\mp}$ |
|---|---|---|---|---|---|---|---|
| $KAl_4^-$ | $Al_4^{2-}$ | 0.2953 | 0.6365 | 0.3413 | 0.0044 | 0.0095 | 0.0051 |
| ($C_{\infty v}$) | $K^+$ | 0.7047 | 0.3635 | -0.3413 | 0.0106 | 0.0054 | -0.0051 |
| $KAl_4^-$ | $Al_4^{2-}$ | 0.1614 | 0.7198 | 0.5584 | 0.0023 | 0.0101 | 0.0078 |
| ($C_{2v}$) | $K^+$ | 0.8386 | 0.2802 | -0.5584 | 0.0118 | 0.0039 | -0.0078 |
| $KAl_4^-$ | $Al_4^{2-}$ | 0.0982 | 0.7597 | 0.6615 | 0.0008 | 0.0066 | 0.0057 |
| ($C_{4v}$) | $K^+$ | 0.9018 | 0.2403 | -0.6615 | 0.0078 | 0.0021 | -0.0057 |

***Table 6.*** Group Fukui Function ($f_g^+, f_g^-$) and Group Philicity ($\omega_g^+, \omega_g^-$) Values for Nucleophilic and Electrophilic Attacks Respectively for the Ionic Units of Different Isomers of $CuAl_4^-$.

| Isomers | Ionic Unit | $f_g^+$ | $f_g^-$ | $\Delta f_g^{\mp}$ | $\omega_g^+$ | $\omega_g^-$ | $\Delta\omega_g^{\mp}$ |
|---|---|---|---|---|---|---|---|
| $CuAl_4^-$ | $Al_4^{2-}$ | 0.6823 | 0.8044 | 0.1221 | 0.0031 | 0.0036 | 0.0006 |
| ($C_{\infty v}$) | $Cu^+$ | 0.3177 | 0.1955 | -0.1221 | 0.0014 | 0.0009 | -0.0006 |
| $CuAl_4^-$ | $Al_4^{2-}$ | -0.1552 | 0.9166 | 1.0718 | 0.0036 | 0.0036 | 0.0048 |
| ($C_{2v}$) | $Cu^+$ | 1.15518 | 0.0834 | -1.0718 | 0.0008 | 0.0008 | -0.0048 |
| $CuAl_4^-$ | $Al_4^{2-}$ | 0.5752 | 1.0738 | 0.4986 | 0.0178 | 0.0332 | 0.0154 |
| ($C_{4v}$) | $Cu^+$ | 0.4248 | -0.0738 | -0.4986 | 0.0131 | -0.0023 | -0.0154 |

***Table 7.*** Group Fukui Function ($f_g^+, f_g^-$) and Group Philicity ($\omega_g^+, \omega_g^-$) Values for Nucleophilic and Electrophilic Attacks Respectively for the $Al_4^{2-}$ Unit of Various Sandwich complexes Based on All-Metal Aromatic Clusters.

| Aromatic Sandwich | Ionic Unit | $f_g^+$ | $f_g^-$ | $\Delta f_g^{\mp}$ | $\omega_g^+$ | $\omega_g^-$ | $\Delta\omega_g^{\mp}$ |
|---|---|---|---|---|---|---|---|
| $Li_2(Al_4TiAl_4)$ | $2Al_4^{2-}$ | 0.8326 | 0.8422 | 0.0096 | 3.0913 | 3.1269 | 0.0356 |
| $Na_2(Al_4TiAl_4)$ | $2Al_4^{2-}$ | 0.7523 | 0.7841 | 0.0318 | 2.6409 | 2.7526 | 0.1116 |
| $K_2(Al_4TiAl_4)$ | $2Al_4^{2-}$ | 0.7024 | 0.7425 | 0.0401 | 2.0784 | 2.1970 | 0.1186 |

***Table 8.*** Energy (E), Polarizability (α), Hardness (η), Chemical potential (μ) and the Electrophilicity (ω) values of Different Isomers of $Al_4^{4-}$, $Li_3Al_4^{-}$ and $Li_4Al_4$ and Sandwich Complexes.

| Molecule | PG | Energy (E) | $\alpha$ | $\eta$ | $\mu$ | $\omega$ |
|---|---|---|---|---|---|---|
| *$Al_4^{4-}$ Isomers* | | | | | | |
| $Al_4^{4-}$ (Singlet) | $D_{\infty h}$ | -969.266 | 1915.578 | 0.954 | 8.062 | 34.066 |
| | $D_{2h}$ | -969.262 | 1910.644 | 1.194 | 8.703 | 31.714 |
| $Al_4^{4-}$ (Triplet) | $D_{\infty h}$ | -969.275 | 2149.025 | 1.206 | 8.719 | 31.521 |
| | $D_{2h}$ | -969.259 | 2276.394 | 1.187 | 8.055 | 27.321 |
| *$Li_3Al_4^{-}$ Isomers* | | | | | | |
| Singlet | $C_S$ | -992.433 | 564.935 | 1.388 | 0.073 | 0.002 |
| Triplet | $C_S$ | -992.434 | 530.666 | 1.460 | 0.003 | 3E-06 |
| Fork | $C_S$ | -992.435 | 522.497 | 1.516 | -0.030 | 2E-04 |
| Hood | $C_2$ | -992.43 | 602.844 | 1.393 | 0.046 | 8E-04 |
| Scooter | $C_1$ | -992.431 | 618.850 | 1.356 | 0.062 | 0.001 |
| Rabbit | $C_{2V}$ | -992.419 | 726.708 | 1.440 | 0.161 | 0.009 |
| *$Li_4Al_4$ Isomers* | | | | | | |
| $C_{2h}$ | | -999.933 | 392.791 | 1.998 | -2.868 | 2.058 |
| $C_{2v}$ | | -999.914 | 364.974 | 1.822 | -3.125 | 2.679 |
| $D_{2h}$ | | -999.932 | 452.724 | 1.920 | -2.579 | 1.732 |
| *Anti-aromatic Sandwich Complexes* | | | | | | |
| $Li_4Al_4Fe(CO)_3$ | | -2603.863 | 370.024 | 2.436 | -3.334 | 2.282 |
| $(Li_4Al_4)_2Ni$ | | -3508.302 | 844.603 | 1.694 | -2.671 | 2.107 |
| bis($Li_4Al_4$nickel(II) chloride) | | -6857.962 | 660.017 | 1.784 | -3.749 | 3.940 |

***Table 9.*** Fukui function ($f^+$, $f^-$) and Philicity ($\omega^+$, $\omega^-$) Values for Nucleophilic and Electrophilic Attacks Respectively for the Aluminum Atoms of Different Isomers of $Al_4^{4-}$(Singlet).

| Isomers | Atom | $f^+$ | $f^-$ | $\omega^+$ | $\omega^-$ |
|---|---|---|---|---|---|
| $Al_4^{4-}$ ($D_{\infty h}$) | Al | 0.7034 | 0.6926 | 23.9627 | 23.5935 |
| | Al | -0.2030 | -0.1930 | -6.9299 | -6.5606 |
| | Al | -0.2030 | -0.1930 | -6.9299 | -6.5606 |
| | Al | 0.7034 | 0.6926 | 23.9627 | 23.5935 |
| $Al_4^{2-}$ ($D_{2h}$) | Al | 0.3913 | 0.4617 | 12.4081 | 14.6425 |
| | Al | 0.1074 | 0.0387 | 3.4070 | 1.2272 |
| | Al | 0.3938 | 0.4609 | 12.4891 | 14.6168 |
| | Al | 0.1075 | 0.0387 | 3.4096 | 1.2274 |

***Table 10.*** Fukui function ($f^+$, $f^-$) and Philicity ($\omega^+, \omega^-$) Values for Nucleophilic and Electrophilic Attacks Respectively for the Aluminum Atoms of Different Isomers of $Al_4^{4-}$(Triplet).

| Isomers | Atom | $f^+$ | $f^-$ | $\omega^+$ | $\omega^-$ |
|---|---|---|---|---|---|
| $Al_4^{4-}$ ($D_{\infty h}$) | Al | 0.7344 | 0.6628 | 20.0640 | 18.1080 |
| | Al | -0.2340 | -0.1628 | -6.4030 | -4.4470 |
| | Al | -0.2340 | -0.1628 | -6.4030 | -4.4470 |
| | Al | 0.7344 | 0.6628 | 20.0640 | 18.1080 |
| $Al_4^{2-}$ ($D_{2h}$) | Al | 0.2492 | 0.2708 | 7.8549 | 8.5349 |
| | Al | 0.2508 | 0.2292 | 7.9056 | 7.2257 |
| | Al | 0.2516 | 0.2285 | 7.9300 | 7.2013 |
| | Al | 0.2484 | 0.2715 | 7.8306 | 8.5593 |

***Table 11.*** Group Fukui Function ($f_g^+, f_g^-$) and Group Philicity ($\omega_g^+, \omega_g^-$) Values for Nucleophilic and Electrophilic Attacks Respectively for the Ionic Units of Different Isomers of $Li_3Al_4^-$.

| $Li_3Al_4^-$ Isomers | Ionic Unit | $f_g^+$ | $f_g^-$ | $\Delta f_g^{\mp}$ | $\omega_g^+$ | $\omega_g^-$ | $\Delta\omega_g^{\mp}$ |
|---|---|---|---|---|---|---|---|
| $Li_3Al_4^-$ ($C_S$ Singlet) | $Al_4^{4-}$ | 0.0974 | 0.5105 | 0.4131 | 0.0002 | 0.0010 | 0.0008 |
| | $Li_3^{3+}$ | 0.9026 | 0.4895 | -0.4131 | 0.0017 | 0.0009 | -0.0008 |
| $Li_3Al_4^-$ ($C_S$ Triplet) | $Al_4^{4-}$ | 0.1316 | 0.4560 | 0.3244 | 4.2E-07 | 1.5E-06 | 1E-06 |
| | $Li_3^{3+}$ | 0.8684 | 0.5440 | -0.3244 | 2.8E-06 | 1.7E-06 | -1E-06 |
| $Li_3Al_4^-$ ($C_S$ Fork) | $Al_4^{4-}$ | 0.0654 | 0.6091 | 0.5437 | 1.4E-05 | 0.0001 | 0.0001 |
| | $Li_3^{3+}$ | 0.9346 | 0.3909 | -0.5437 | 0.0002 | 8.6E-05 | -0.0001 |
| $Li_3Al_4^-$ ($C_2$ Hood) | $Al_4^{4-}$ | 0.0632 | 0.4963 | 0.4331 | 4.9E-05 | 0.0004 | 0.0003 |
| | $Li_3^{3+}$ | 0.9368 | 0.5037 | -0.4331 | 0.0007 | 0.0004 | -0.0003 |
| $Li_3Al_4^-$ ($C_1$ Scooter) | $Al_4^{4-}$ | -0.2498 | 0.4715 | 0.7213 | -0.0004 | 0.0007 | 0.0010 |
| | $Li_3^{3+}$ | 1.2498 | 0.5285 | -0.7213 | 0.0018 | 0.0008 | -0.0010 |
| $Li_3Al_4^-$ ($C_{2v}$ Rabbit) | $Al_4^{4-}$ | 0.0146 | 0.5695 | 0.5549 | 0.0001 | 0.0051 | 0.0050 |
| | $Li_3^{3+}$ | 0.9854 | 0.4305 | -0.5549 | 0.0088 | 0.0039 | -0.0050 |

***Table 12.*** Group Fukui Function ($f_g^+, f_g^-$) and Group Philicity ($\omega_g^+, \omega_g^-$) Values for Nucleophilic and Electrophilic Attacks Respectively for the Ionic Units of Different Isomers of $Li_4Al_4$.

| $Li_4Al_4$ Isomers | Ionic Unit | $f_g^+$ | $f_g^-$ | $\Delta f_g^{\mp}$ | $\omega_g^+$ | $\omega_g^-$ | $\Delta\omega_g^{\mp}$ |
|---|---|---|---|---|---|---|---|
| $Li_4Al_4$ ($C_{2h}$) | $Al_4^{4-}$ | 0.3053 | 0.5018 | 0.1965 | 0.6283 | 1.0328 | 0.4044 |
| | $Li_4^{4+}$ | 0.6947 | 0.4982 | -0.1965 | 1.4298 | 1.0254 | -0.4044 |
| $Li_4Al_4$ ($C_{2v}$) | $Al_4^{4-}$ | 0.3420 | 0.6032 | 0.2612 | 1.4791 | 1.0016 | 0.7000 |
| | $Li_4^{4+}$ | 0.6580 | 0.3968 | -0.2612 | 1.2003 | 1.6777 | -0.7000 |
| $Li_4Al_4$ ($D_{2h}$) | $Al_4^{4-}$ | -0.0220 | 0.7314 | 0.7532 | -0.0377 | 1.2669 | 1.3045 |
| | $Li_4^{4+}$ | 1.0217 | 0.2686 | -0.7532 | 1.7697 | 0.4652 | -1.3045 |

***Table 13.*** Group Fukui Function ($f_g^+, f_g^-$) and Group Philicity ($\omega_g^+, \omega_g^-$) Values for Nucleophilic and Electrophilic Attacks Respectively for the $Al_4^{2-}$ Unit of Various Sandwich complexes Based on All-Metal Anti-Aromatic Clusters.

| Anti-aromatic Sandwich | Ionic Unit | $f_g^+$ | $f_g^-$ | $\Delta f_g^{\mp}$ | $\omega_g^+$ | $\omega_g^-$ | $\Delta\omega_g^{\mp}$ |
|---|---|---|---|---|---|---|---|
| $Li_4Al_4Fe(CO)_3$ | $Al_4^{4-}$ | 0.2400 | 0.3820 | 0.1420 | 0.5476 | 0.8717 | 0.3241 |
| $(Li_4Al_4)_2Ni$ | $2Al_4^{4-}$ | 0.2648 | 0.4810 | 0.2162 | 0.5578 | 1.0133 | 0.4555 |
| bis($Li_4Al_4$ nickel(II) chloride) | $2Al_4^{4-}$ | 0.8425 | 0.8471 | 0.0046 | 3.3193 | 3.3375 | 0.0183 |

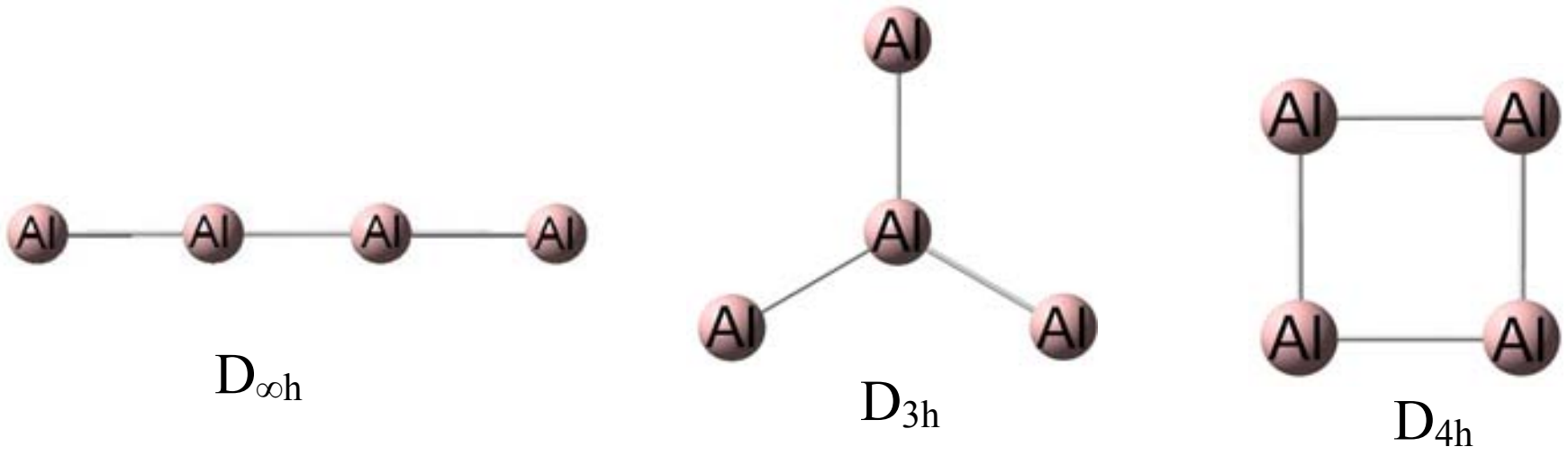

***Figure 1.*** Optimized structures of various isomers of $Al_4^{2-}$.

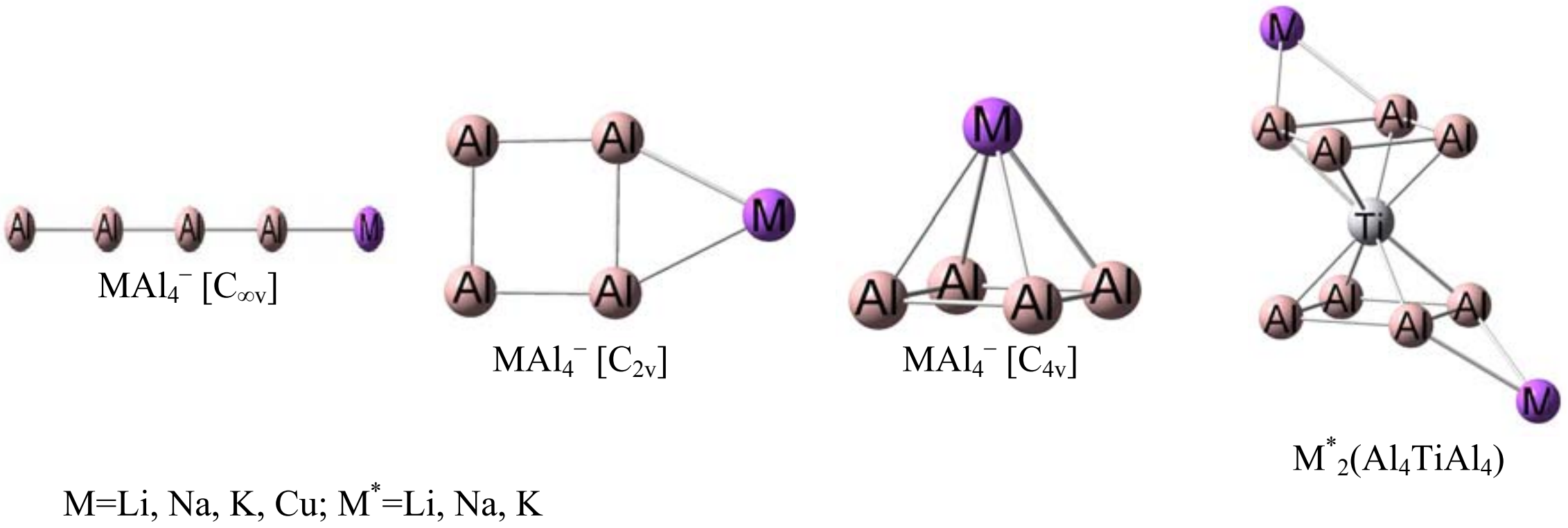

***Figure 2.*** Optimized structures of various isomers of $MAl_4^-$ (M ≡ Li, Na, K, Cu) and aromatic sandwich complexes $M^*_2(Al_4TiAl_4)$ ($M^*$ ≡ Li, Na, K).

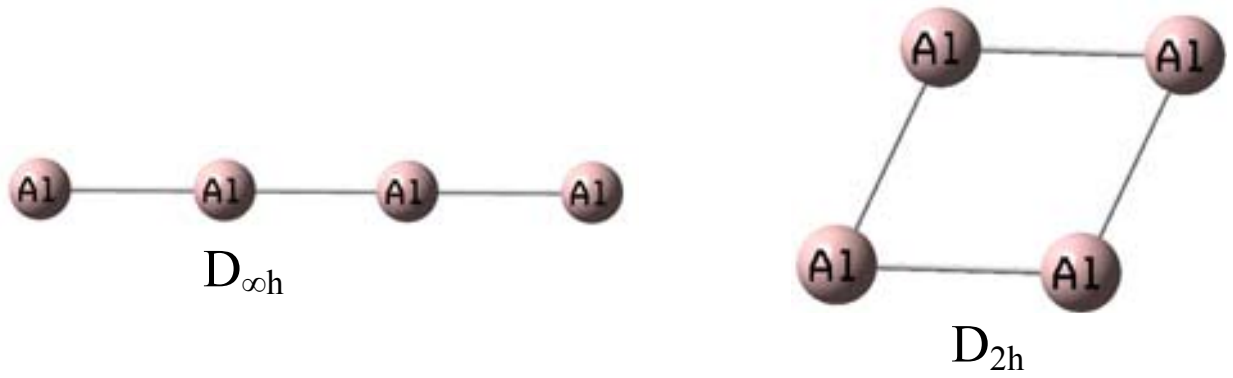

***Figure 3.*** Optimized structures of isomers of $Al_4^{4-}$.

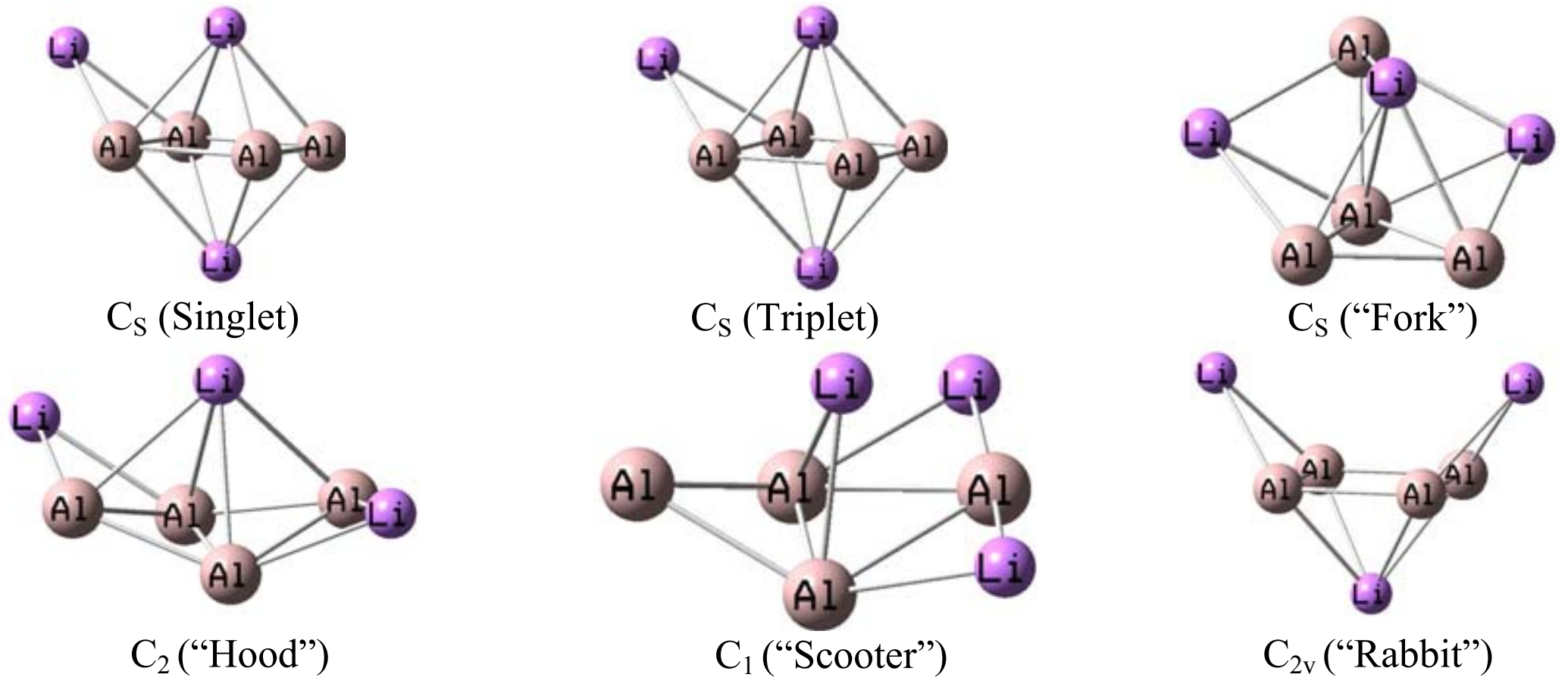

***Figure 4.*** Optimized structures of isomers $Li_3Al_4^-$.

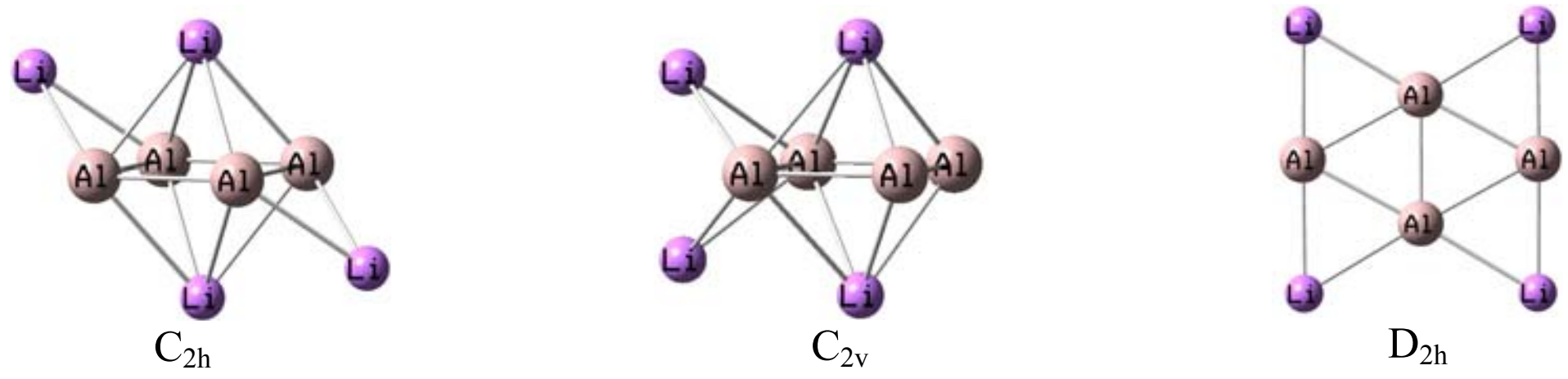

***Figure 5.*** Optimized structures of isomers of $Li_4Al_4$.

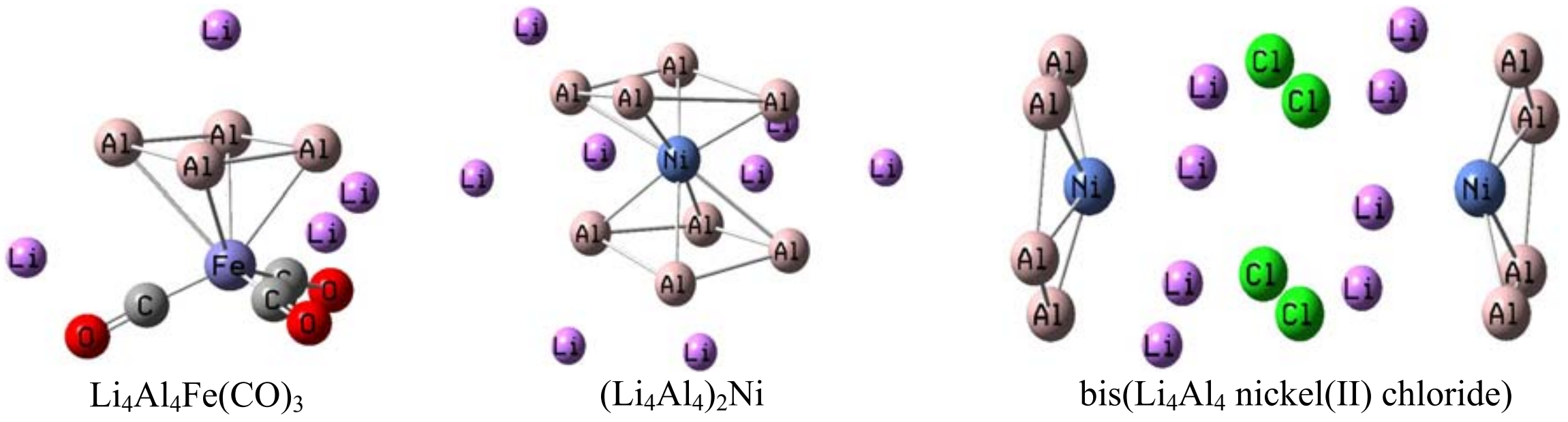

***Figure 6.*** Optimized structures of the sandwich complexes of $Li_4Al_4$.